\documentclass{article}
\usepackage[utf8]{inputenc}
\usepackage{hyperref}
\usepackage{amsmath}
\usepackage{amssymb}
\usepackage{booktabs}
\usepackage{adjustbox}
\usepackage{graphicx}
\usepackage{authblk}
\usepackage{float}
\usepackage{tabularx}
\usepackage[margin=2.8cm]{geometry}
\title{Large sensory analysis of vegetables from  conventional, organic and no-till practices}
\date{October 2022}

\author[1]{S. Loustau}
\author[1]{F. Lefer}
\author[2]{S. Ducos}

\affil[1]{Université de Pau et des Pays de l'Adour, Laboratoire de Mathématiques et de leurs Applications, France}
\affil[2]{Université de Pau et des Pays de l'Adour, Institut des Sciences Analytiques et de Physico-Chimie pour l'Environnement et les Matériaux, France}

\begin{document}
    
\maketitle
\begin{abstract}
   There is a growing interest in agriculture to address soil deterioration issues and achieve a sustainable agrosystem. Many producers introduce no-till techniques and show significant yields with better ecosystem functioning, as well as improved carbon and water cycles. However, sensory analysis to compare vegetables from these practices are scarce. In this paper, we conduct triangle and hedonic tests over 15 panels of consumers for a total of more than 950 consumers. Based on statistical hypothesis testing and maximum likelihood estimation, significant statistical differences (P $< 0.05$ and less) are produced for both triangle tests and hedonic ranking in some specific cases. These results show a trend for sensory preferences of no-till practices.
   \end{abstract}

%\keyword{conservation agriculture; no-till; sensory analysis; statistics; vegetables; triangle test; hedonic test; panel} 

\section{Introduction}
\label{sec:intro}
%agri-food, soil and consumers
%Conservation agriculture (CA) is characterised by s
Sustainable agricultural practices are motivated by the convergence of Agriculture Science and Ecology. Many researchers in soil sciences, agronomy and ecosystems have studied the impact of biodiversity in the future of the ecosphere (see \cite{19}). High diversity mixtures activate complementarity, decrease diseases and improve nutrient-cycling feedbacks and retention over the long term. On the contrary, biodiversity loss strongly impacts ecosystem functioning \cite{20}. In soil sciences, if biodiversity is kept at an optimum, ecosystem services including edible and nonedible biological products, provision of clean water, hydrologic and nutrient cycling, habitats for microorganisms and mesofauna, and soil organic matter decomposition are preserved. Moreover carbon sequestration and climate regulation can be warranted  in the context of global carbon cycling models (see \cite{10}). As a result, soil management in agricultural landscapes should deploy production practices that are in harmony with soil-mediated organisms that deliver these broad range of direct and indirect positive benefits to human well-being, and have an impact on our survival and quality of life. However, despite a growing interest and an empirical demonstration of the transformation of agricultural production (see \cite{23}), conventional agri-food systems is firmly established. Driven by a remarkable growth of calorie production per capita, this substancial success has come at a great cost \cite{21}. It contributed to social and environmental problems, and resulted in physical and chemical soil degradation, as well as soil components transportation by water and wind erosion \cite{11}.  Nowadays, this industrial approach to food is among the major threats to global sustainability (see \cite{22}).

%market gardening, no-till precisions
Regarding these challenges, the Food and Agriculture Organization (FAO) has defined Conservation Agriculture (CA) as a range of practices based on three main principles: (1) minimum soil disturbance obtained using reduced-tillage or no-tillage; (2) a permanent soil cover via living or dead (mulch) cover crops; and (3) diversified crop rotations (see \cite{27}). Indeed, it is now widely known that reducing tillage can offer a number of economic and environmental advantages for the preparation of soils to crop establishment. \cite{26} reviews a large literature enumerating pros and cons of soil tillage versus no-till practices, based on an interdisciplinary large research program conducted from 1980 to 2000. Despite many agricultural and practical pros, tillage causes damages on soil fertility, like soil erosion by water and wind, soil sealing and crusting of the soil surface have been proved, as well as deep soil compaction, reduced organic matter content of the soil, and then high emission rates of CO2 through fast biological oxidation of organic matter. Moreover, damages on drinking water, by leaching and run-off of nitrate, water soluble phosphate and other xenobiotica contributes to the overall environmental problem. 

In this paper, we are interested in particular market gardening practices based on CA practices, low mechanization and no-till.  Inspired by \cite{25}, no-till (also known as direct drilling and zero tillage) can be defined as an agrosystem in which crops are sown without any prior loosening of the soil by cultivation other than the very shallow disturbance ($< 5$ cm) which may arise by the passage of the drill coulters and after which the surface remains always covered with plant residues or additional organic matter. In market gardening, these techniques are associated with an alternative to the standard approach of agricultural productivity, also called micro-farm model, where the farmer provides a sustainable land management by reducing the size of its land use, and developping handmade practices and minimum soil disturbance (see \cite{morel2017}). Crops are planted in permanent raised beds, that is raised mounds of earth about 10-15 centimetres (or roughly 4-6 inches) from the ground, separated by alleys to allow for easy circulation. Cover crops and roller crimpers are used instead of pesticides and tillers. These techniques change significantly the job of market gardener, its finance and access to capital.  It ensures to restore many important soil properties, such as structure promoting efficient drainage and aeration, and minimizes loss of topsoil via erosion. This quality is mainly due to the significant rate of soil organic matter (see \cite{28}), and illustrates the Inverse Productivity (IP) relationship (see \cite{9}), showing that output per unit of land tends to be greater on these small farms. It shows that rebuilding soils is then necessary to arrive at a resilient and productive agriculture, in order to increase yields while minimizing environmental harm.
% conclusion plutôt ? It is now admitted for a long time that an inverse relationship occurs between farm size and yield. It has been observed on ICRISAT data (International Crops Research Institute for the Semi-Arid Tropics, see https://www.fao.org/3/i5251e/i5251e.pdf) and also in developing countries (see \cite{9}). This phenomenon, called the Inverse Productivity (IP) relationship shows that output per unit of land tends to be greater on small farms than on large farms. 

%consumer and market, organic products In the history, the role of the consumer has changed. Limited to voting with their wallets, the role of the consumer https://www.mdpi.com/2077-0472/12/2/203#B23-agriculture-12-00203 

% Flo version 
%ok dans l'ensemble bon travail
%le debut manque de ref (trop de phrases sans appuis scientifique, j'ai mis deux zones importantes ou manque de ref, il faut que tu en trouves des pertinentes)
%j'ai ajoute la partie food comparison till vs no-till a la fin
From the food quality point of view, there is a growing demand for products that are free from pesticides and environmentally sustainable \cite{29}. Indeed, the role of the consumer has evolved. Instead of just voting with their wallets, consumers now have a stronger influence. They care more about where their food comes from and how it affects environment and society \cite{30, 29}. As a result, Consumers are eager to reconnect with the entire process of food production, including the farmers and producers involved. This shift in consumer preferences has driven the expansion of organic markets \cite{31, 32}. These markets offer products that are perceived to be of higher quality and healthier. However, when it comes to scientific research on the nutritional value and health benefits of organic foods, there is some conflicting results. With respect to product quality, several papers attest significant differences in nutrients and mineral contents between organic vs conventional vegetables (potatoes and corn, see \cite{6}). Another study compared nutritive quality of potatoes from organic farms to the conventional tillage counterpart in \cite{7}. Numerous comparative studies between organic and conventional vegetables have consistently demonstrated superior nutritive qualities in organic crops. These qualities include higher vitamin C content and significantly lower nitrate levels \cite{Rembialkowska, Theuer, Worthington_1998}, which contribute to a double anti-carcinogenic impact. However, another review \cite{15} compiles several research results and claims that although there are some differences in nutrients content between organic and conventional agriculture, it remains difficult to prove that organic food improves human well-being or health after consumption. Indeed, despite some systematic differences in the nutritional content, i.e. nitrogen, protein, vitamin C, and nitrates, no clear effect on health-related biomarkers could be established due to the difficulty of conducting nutritional tests in humans, where all factors influencing human health must be kept constant. Brandt and Mølgaard \cite{Brandt} reached the conclusion that significant nutritional disparities between organic and conventional feed are improbable, given the absence of deficiencies in vitamins, proteins, minerals, and carbohydrates in "typical First World diets." They also highlighted that the presence of pesticide residues observed in conventional crops should not be a cause for concern. Nonetheless, they note that secondary metabolites in the diet tend to be below optimal levels. Considering that organic vegetables and fruits exhibit higher proportions of these secondary metabolites compared to their conventional counterparts, there is a potential for greater benefits to human health. Finally, conventional till and no-till techniques have been recently compared. In \cite{12}, two varieties of Tunisian wheat are studied from the raw grain to the end product (spaghetti) under two different soil management systems (namely conventional and long-term CA) over two years. The grain quality, the dough texture and the pasta quality (yellow index) are measured whereas in \cite{13}, the effect of tillage and monocropping over grain yield and quality is investigated. Conservation Agriculture and rotation demonstrated significant influences on grain quality because the inclusion of rotation favored higher N-remobilization to the grains.

%sensorial analysis for no-till
From the sensory analysis point of view, studies comparing vegetables from these techniques are scarce, and there are still some limitations to claim that no-till techniques and smallholders food is better than organic or conventional. In \cite{17}, a comparison of conventional and organically lettuces (as well as hydropony) was performed, and no significant difference has been proved. However, as claimed in \cite{18}, many factors need to be paired in each production technique to have a rigorous and comparable produce quality study. In this paper, to attack this challenge, we conduct a large sensory evaluation over 15 panels made of a total of 942 untrained consumers in order to test the ability to distinguish conventional, organic and conservation techniques over 7 vegetables. For that purpose, we use standard techniques from sensory analysis, namely triangle and hedonic test, coupled with a rigorous statistical analysis mixing test theory and maximum likelihood estimation on a big dataset of around 1000 consumers. 

The paper is organized as follows. In Section 2, we describe the sensory analysis and the experiences we conduct. In Section 3, we describe the statistical model and analysis whereas Section 4 presents the main quantitative results. Section 5 concludes with a summary of the contribution and a discussion for future works.

%The introduction should briefly place the study in a broad context and highlight why it is important. It should define the purpose of the work and its significance. The current state of the research field should be reviewed carefully and key publications cited. Please highlight controversial and diverging hypotheses when necessary. Finally, briefly mention the main aim of the work and highlight the principal conclusions. As far as possible, please keep the introduction comprehensible to scientists outside your particular field of research. Citing a journal paper \cite{ref-journal}. Now citing a book reference \cite{ref-book1,ref-book2} or other reference types \cite{ref-unpublish,ref-communication,ref-proceeding}. Please use the command \citep{ref-thesis,ref-url} for the following MDPI journals, which use author--date citation: Administrative Sciences, Arts, Econometrics, Economies, Genealogy, Humanities, IJFS, Journal of Intelligence, Journalism and Media, JRFM, Languages, Laws, Religions, Risks, Social Sciences, Literature.
%%%%%%%%%%%%%%%%%%%%%%%%%%%%%%%%%%%%%%%%%%
\section{Materials and Methods}
%Materials and Methods should be described with sufficient details to allow others to replicate and build on published results. Please note that publication of your manuscript implicates that you must make all materials, data, computer code, and protocols associated with the publication available to readers. Please disclose at the submission stage any restrictions on the availability of materials or information. New methods and protocols should be described in detail while well-established methods can be briefly described and appropriately cited.

%Research manuscripts reporting large datasets that are deposited in a publicly avail-able database should specify where the data have been deposited and provide the relevant accession numbers. If the accession numbers have not yet been obtained at the time of submission, please state that they will be provided during review. They must be provided prior to publication.

%Interventionary studies involving animals or humans, and other studies require ethical approval must list the authority that provided approval and the corresponding ethical approval code.
\subsection{Sensory analysis context}
Sensory evaluation tries to isolate the sensory properties of foods and other products and can be defined as a scientific method that is used to evoke, measure and interpret responses to the sensory properties through the senses \cite{1}. Sensory evaluation assesses these properties, such as flavour, consistency, texture, using trained panelists or untrained consumer groups for the purpose of rating the quality of a product, or comparing one product to another \cite{3}. In this paper, we build  untrained consumer groups based on 13 different experiences realized under the protocol described below. The vegetables included in the study were also chosen based on available Bearn production. 
\subsection{Experience protocol}
\subsubsection{Sources of vegetables}
Seven fruits and vegetables were tested during the study, namely fava beans, green peas, zucchinis, beetroots, cucumbers, tomates and peppers. Of the 30 sample batches purchased for the study, 22 were sourced from local farm in Bearn, whereas 5 from local organic store (see Table \ref{table_1} below). We omit large supermarkets to reduce variations in freshness due to a larger distribution channel, as well as unknown exact origin of the product (2 samples in the study). The study took place in a reduced range of time, from June to September, 2023, in a small area around the Pau agglomeration, in Béarn, France, to reduce variability due to climate, temperatures, or water disponibility for the plants. The vegetables compared for each event belong to the same variety. Table \ref{table_1} compiles the variety of vegetables and producers of our samples, its origins, for each event and each vegetable, and the associated agronomic practices (no-till, organic or conventional). 
\subsubsection{Sample preparation}
Sample batches of each food were purchased in the morning of the analysis, or the day before, to have fresh samples (vegetables from direct producers were bought before vegetables from retails to have approximately equal degrees of freshness). Preparation was conducted for each sample batch a few hours before the event and depended on the nature of the samples. Table \ref{table_2} shows the preparation associated to each vegetable.

\begin{table}[H]
\newcolumntype{C}{>{\centering\arraybackslash}X}
\begin{tabularx}{\textwidth}{C|CCCCCC}
\toprule
\textbf{Event id} & \textbf{Date} & \textbf{Vegetable} & \textbf{Test} & \textbf{Origin 1} & \textbf{Origin 2} & \textbf{Sample size} \\
\hline
\midrule
1 & 06/01 & Fava bean & NT/Org & Farm\textsuperscript{*} & Organic store\textsuperscript{§} & 54 \\
2 & 06/08 & Green pea & NT/Org & Farm\textsuperscript{*} & Organic store\textsuperscript{§} & 45 \\
3 & 06/08 & Zucchini Black beauty & Org/Conv & Farm\textsuperscript{*} & Local products Store\textsuperscript{*} & 43\\
4 & 06/15 & Zucchini Black beauty & NT/Conv & Farm\textsuperscript{*} & Farm \textsuperscript{*} & 85 \\
5 & 06/16 & Beetroot Detroit dark red & Org/Conv & Organic Store\textsuperscript{†} & Local products Store\textsuperscript{†} & 47 \\
6 & 06/22 & Cucumber Noa & NT/Conv & Farm\textsuperscript{*} & Farm\textsuperscript{*} & 45 \\
7 & 06/22 & Beetroot Detroit dark red & NT/Org & Farm\textsuperscript{*} & Organic store\textsuperscript{*} & 47 \\
8 & 06/30 & Cucumber Noa & NT/Org & Farm\textsuperscript{*} & Organic store\textsuperscript{*} & 95 \\
9 & 07/06 & Tomato Black Krim & NT/Org & Farm\textsuperscript{*} & Farm\textsuperscript{*} & 71 \\
10 & 07/07 & Tomato Paola & Org/Conv & Farm\textsuperscript{*} & Farm\textsuperscript{*} & 88 \\
11 & 07/11 & Tomato Oxheart & NT/Org & Farm\textsuperscript{*} & Farm\textsuperscript{*} & 25 \\
12 & 07/12 & Tomato Rose de Berne & Org/Conv & Farm\textsuperscript{*} & Farm\textsuperscript{*} & 79 \\
13 & 07/21 & Pepper Bull horn & Org/Conv & Farm\textsuperscript{*} & Farm\textsuperscript{*} & 102 \\
14 & 09/02 & Tomato Paola & NT/Conv & Farm\textsuperscript{*} & Farm\textsuperscript{*} & 25 \\
15 & 09/07 & Tomato Black Krim & NT/Conv & Farm\textsuperscript{*} & Farm\textsuperscript{†} & 64 \\

\bottomrule
\end{tabularx}
\footnotesize\textsuperscript{*}Bearn. \textsuperscript{†}Occitanie. \textsuperscript{§}Unknown.
\caption{Events description: origin and sample sizes.}
\label{table_1}
\end{table}

\begin{table}[H]
\label{tab1}
\newcolumntype{C}{>{\centering\arraybackslash}X}
\begin{tabularx}{\textwidth}{>{\hsize=0.3\hsize}X>{\hsize=0.7\hsize}X}
\toprule
\textbf{Vegetable} & \textbf{Method}\\
\hline
\midrule
Fava bean & Peeled \\
Green pea & Roughly mixed \\
Zucchini & Ends removed, skin on, sliced  \\
Cucumber & Ends removed, skin on, sliced and diced in quarters  \\
Beetroot & Peeled, sliced and diced  \\
Tomato & Stem and tough part around the stem removed, cut in wedges  \\
Pepper & Ends removed, sliced  \\
\bottomrule
\end{tabularx}
\caption{Food Preparation by Vegetable}
\label{table_2}
\end{table}

\subsubsection{Panel selection and data collection}
\begin{figure}
    \centering
    \includegraphics[width=10cm]{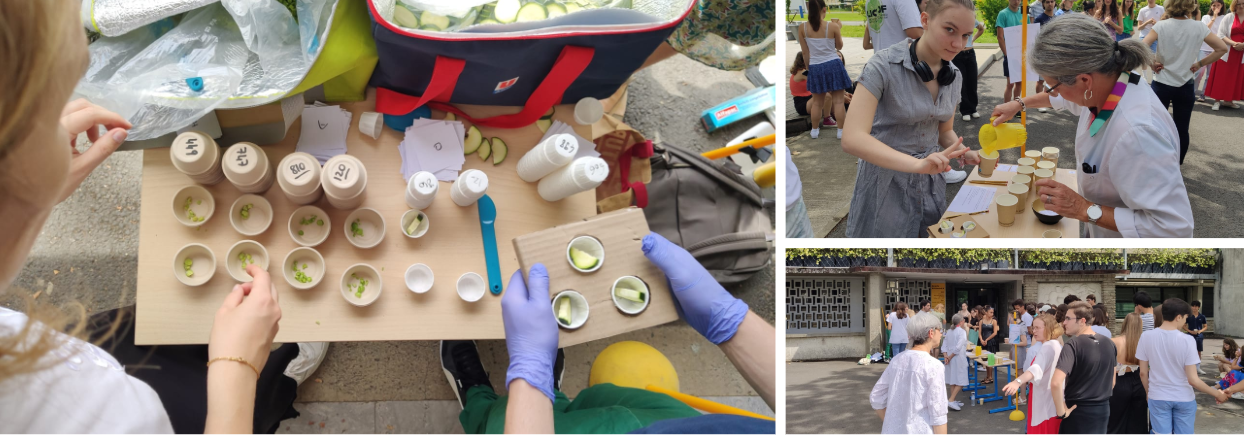}
    \caption{Preparation of the test - event in Lycée Louis Barthou (event id 2 and 3)}
    \label{fig1}
\end{figure}

The sensory panel was recruited for each experience, and took place before lunch. The criteria for panelists were the following: they were available to volunteer for the experience, smokers and nonsmokers. The proportion of smokers in the study was rather small (9,9\% in the whole dataset). Each volunteer that has consumed food, coffee or tea, or has smoked the hour before has been excluded from the study. We add in front of the samples, mineral water as a palate cleanser between tastings, and instructions for the panelists in order to respect the standardized protocol of tasting procedure (see Figure \ref{fig1}). The whole protocol works as follows.
\begin{enumerate}
\item \textbf{Before the Test :}
\begin{itemize}
\item No smoking, eating, chewing gum, or drinking coffee during the hour prior to the test. 
\item The panelist has no information on the purpose of the test nor on the origin of the samples.
\end{itemize}
\item  \textbf{During the Test:}
\begin{itemize}
\item The panelist is presented with three samples arranged in a triangle shape, the combinations being randomized across the panel. 
\item The panelist can taste the samples in any order, and can taste each sample more than once.
\item The panelist was not allowed to ask for more food in order to make a response.
\item The panelist must rinse their palate between each sample.
\item The panelist must not talk before they are finished.
\end{itemize}
\item  \textbf{After the Test:}
The panelist completes an answer sheet with the number of the sample they think is the odd one,
their preference between the odd one and the other 2, and finally if they smoke or not.
\end{enumerate}

%%%%%%%%%%%%%%%%%%%%%%%%%%%%%%%%%%%%%%%%%%
\section{Statistical model and analysis}
\label{sec:stat}
Triangle test is the most used technique in sensory analysis to compare different products. This approach is usually based on a statistical test. In what follows, we propose an hybrid testing and estimating procedure in order to add insights into the number of correct answers not due to chance in the triangle test, and derive a suitable preference score for each product in each experiment. The model described below in \eqref{model} supposes the existence of a proportion $p\in [0,1]$ of judges abble to detect the difference between the two products. This model is largely adopted in sensory analysis even if it is not rigorously specified in many papers. It ensures a realistic hypothesis over the sample, that is a non-iid set of observations over the population, since each consumer have its own capacity to detect (or not) a difference. With this model at hand, the statistical analysis allows to reject rigorously, for each experience, the null hypothesis "$H_0:p=0$", and estimate the preferences of the population of each experience. 
\subsection{Statistical model}
We suppose we have at hand a sample of independent binary random variables $\{X_1, \ldots, X_n\}$, where $X_i=1$ when judge $1\leq i\leq n$ has a correct answer (and $0$ otherwise). The law of $X_i,\, i=1, \ldots n$ depends on judge $i=1, \ldots, n$ as follows. When judge $i$ is not able to detect, the law of $X_i$ is a Bernoulli distribution $\rho\sim \mathcal{B}\left(\frac{1}{3}\right)$ (purely random choice) whereas when judge $i$ is able to detect, $X_i$ has law $\nu\sim\mathcal{B}(1)$ (correct answer). The statistical problem is to decide if the number $m\leq n$ of observations based on distribution $\nu$ is non-zero, based on the classical statistic:
\begin{equation}
\label{model}
S_n = \sum_{i=1}^nX_i.
\end{equation}
It is important to note that $S_n$ is a binomial random variable only under the null hypothesis "$H_0:m = 0$", where in this case, each $X_i\sim \rho$ and the standard iid assumption is satisfied. Then, the classical procedure of triangle test, based on the distribution of $S_n$ under the null hypothesis, can reject $H_0$ with a given probability - or risk -  $0<\alpha<1$ with a straightforward binomial test.
\subsection{Maximum likelihood estimation}
Based on $S_n$, we can also estimate the number of correct answers not due to chance as follows. Thanks to the model we have at hand, the number of correct answers $S_n$ in \eqref{model} can be rewritten as:
\begin{equation}
\label{model2}
S_n = m + Y,
\end{equation}
where $m$ denotes the true - and unknown - number of judges abble to detect a difference, and $Y\sim\mathcal{B}\left(n-m,\frac{1}{3}\right)$ is the number of correct answers due to chance. With the previous testing approach, if $m=0$, we are under the null hypothesis, where all the correct answers are due to chance. Here we propose to estimate the number of correct answers not due to chance based on the observation of $S_n$. For that purpose, we propose a maximum likelihood estimator $\hat{k}$ of the correct answers due to chance defined as follows:
\begin{equation}
\label{mle}
\hat{k}=\arg\max_{k=0, \ldots ,n}\mathbb{P}(S_n|Y = k),
\end{equation}
where $\mathbb{P}(S_n|Y=k)$ denotes the conditional distribution of the observations, when $k$ is the actual number of correct answers due to chance. Estimator $\hat{k}$ is then the most probable number of correct answers due to chance, given the observations. Then, thanks to \eqref{model2}, we estimate the correct answers not due to chance by:
\begin{equation}
\label{mlefinal}
\hat{m}=S_n-\hat{k}.
\end{equation}
We can prove that \eqref{mlefinal} is a biased estimator. However, as shown in \cite{24}, the bias becomes neglictable when the number of observations exceeds $n=20$. Finally, with \eqref{mlefinal}, we propose in Section \ref{sec:results} to add an estimation of the preferences for each product, based on the estimation and the MLE procedure described above. For that purpose, we substract the correct answers due to chance from the observed preferences of the whole correct answers, in order to give a more realistic analysis.
\section{Results}
\label{sec:results}
The results of this paper are based on a panel of 942 consumers, hired from 15 experiences with 8 different vegetables realized from June to September, 2023 in Bearn, France. As discussed in Section \ref{sec:intro}, three different agricultural practices are tested in the study, namely conventional (Conv), organic (Org) and no-till (NT). All the results are based on the two-step procedure described in Section \ref{sec:stat}. A triangle test allows to show if some panelists of the sample are able to detect a difference between two vegetables, whereas an hedonic ranking reveals the tasting preferences of the consumers, for each experience. In this section, we report the main results experience by experience, and then proposes some overall statistics and analyses, based on the whole sample.
\subsection{Results event by event}
In Table \ref{table_4} below we detail the results event by event. In these 15 experiences, sample sizes vary from 25 to 102 consumers. Each consumer has tasted two vegetables of the same variety with comparable origin (see Table \ref{table_1}) and according to two agricultural methods among conventional (Conv), organic (Org) or no-till (NT). For each experience, we compute a p-value for the triangle test to reject the null hypothesis "$H_0:m=0$" in model \eqref{model2}, namely the number of consumers able to detect a difference is non-zero. Table \ref{table_4} illustrates a majority (10 out of 15) of significant triangle tests (P$<0.05$). Biggest p-values ranges around $10\%$ (event 1 and 13), except event 14 with $0.46$, whereas smallest values are around zero (event 8 and event 11, and event 15). It shows that in general, we can claim that some consumers are able to detect a difference, with a risk lower than 5$\%$. This result is confirmed by the estimated correct answers not due to chance, based on \eqref{mle}, with values from 1 (the minimum for event 14) to 40 consumers (the maximum, for event 15). Then, we show the preferences of the consumer groups, based on correct answers (442 among 942 initial responses).The preferred type of vegetables is dominated by NT and Org but unlike triangle tests statistics, a quick look at the p-values of the hedonic test reveals really less significant results. 8 out of 15 p-values are bigger than $25\%$ (among then, 5 are bigger than $40\%$). However, 4 of them are significant (P$<0.05$), but among them, event 7 and event 9 show contradictory results between NT and Org. 

\begin{table}[H]
\newcolumntype{C}{>{\centering\arraybackslash}X}
\begin{tabularx}{\textwidth}{C|CCCCCCCC}
\toprule
\textbf{Event id} & \textbf{Test} & \textbf{Sample size} & \textbf{Correct answers} & \textbf{p-value triangle test} & \textbf{$\hat{m}$ \eqref{mlefinal}} & \textbf{Preferred type} & \textbf{p-value hedonic test} & \textbf{Est. preferences \eqref{mlefinal}} \\
\hline
\midrule
1 & NT/Org & 54 & 22 & 0.156 & 7 & NT & 0.416 & 4.5 / 2.5 \\
2 & NT/Org & 45 & 21 & \textbf{0.043} & 9 & NT & 0.5 & 5 / 4 \\
3 & Org/Conv & 43 & 23 & \textbf{0.005} & 13 & Conv & 0.5 & 6 / 7 \\
4 & NT/Conv & 85 & 35 & 0.08 & 10 & NT & \textbf{0.044} & 10 / 0 \\
5 & Org/Conv & 47 & 24 & \textbf{0.009} & 13 & Org & 0.27 & 8.5 / 4.5 \\
6 & NT/Conv & 45 & 22 & \textbf{0.021} & 11 & Conv & 0.066 & 1.5 / 9.5 \\
7 & NT/Org & 47 & 24 & \textbf{0.009} & 13 & Org & \textbf{0.032} & 1.5 / 11.5 \\
8 & NT/Org & 95 & 47 & \textbf{0.0008} & 23 & NT & 0.28 & 14 / 7 \\
9 & NT/Org & 71 & 32 & \textbf{0.026} & 13 & NT & \textbf{0.035} & 11.5 / 0.5 \\
10 & Org/Conv & 88 & 36 & 0.083 & 10 & Org & 0.121 & 9 / 0 \\
11 & NT/Org & 25 & 17 & \textbf{0.0004} & 14 & NT & 0.166 & 9.5 / 4.5 \\
12 & Org/Conv & 79 & 37 & \textbf{0.009} & 16 & Org & 0.37 & 9.5 / 6.5 \\
13 & Org/Conv & 102 & 40 & 0.125 & 9 & None & 0.5 & 4.5 / 4.5 \\
14 & NT/Conv & 25 & 9 & 0.46 & 1 & NT & 0.5 & 1 / 0 \\
15 & NT/Conv & 64 & 48 & \textbf{0.0000} & 40 & NT & \textbf{0.0000} & 36 / 4 \\
\bottomrule
\end{tabularx}
\footnotesize{\textsuperscript{1}Some testers did not write their preference, which results in missing preference details for events 8, 9, and 10.}
\caption{Results event by event. Bold p-values are significant (p $< 0.05$).}
\label{table_4}
\end{table}
Event 15, where a variety of heirloom tomatoe (Black Krim) was tested, is particular and shows the most powerful differences. More generally, tomatoes show interesting results and are studied separately in what follows.
%This section may be divided by subheadings. It should provide a concise and precise description of the experimental results, their interpretation as well as the experimental conclusions that can be drawn.
\subsection{Aggregation of data and statistics}
From the statistical point of view, the model described in Section \ref{sec:stat} is robust to aggregation of data. Indeed, the non-iid assumption in the sample allows to perform triangle - as well as hedonic test - for the whole dataset. We then propose in the sequel to compile different aggregations of data over the whole dataset. We  derive a vegetable specific study of tomatoes, as well as a global comparison of the three modality Conv, Org an NT, in order to benefit from the repetitiveness of our experiments and the large panel of our study. Table \ref{table_5} summarizes the results of the triangle test and the hedonic ranking comparison by comparison, based on the aggregation of the whole dataset. As highlighted above in Table \ref{table_4}, all p-values for triangles tests are significant in Table \ref{table_5}. Interestingly, for the hedonic test, a significant p-value is observed when we compare NT versus Org tomatoes, as well as NT versus Conv for both the whole dataset set and the specific tomatoe study.

\begin{table}[H]
\newcolumntype{C}{>{\centering\arraybackslash}X}
\begin{tabularx}{\textwidth}{C|CCCCCC}
\toprule
\textbf{Test} & \textbf{Vegetable} & \textbf{Sample size} & Correct answers & \textbf{p-value triangle test} &  \textbf{Preferred type}& \textbf{p-value hedonic test}\\
\midrule
\hline
NT/Org & All & 312 & 146 & \textbf{0.000001} & NT & 0.202 \\
NT/Org & Tomatoes & 71 & 32 & \textbf{0.0262} & NT & \textbf{0.035} \\
Org/Conv & All & 359  & 160 & \textbf{0.000006} & Org & 0.133\\
Org/Conv & Tomatoes & 167 & 73 & \textbf{0.0033} & Org  & 0.097\\
NT/Conv & All & 244  & 131 & \textbf{0.000000} & NT & \textbf{0.000215}\\
NT/Conv & Tomatoes & 114 & 74 & \textbf{0.000000} & NT & \textbf{0.000006} \\
\bottomrule
\end{tabularx}
\caption{Summary of the results - whole study and tomatoes case study}
\label{table_5}
\end{table}

\begin{figure}[H]
\centering
\includegraphics[width=14 cm]{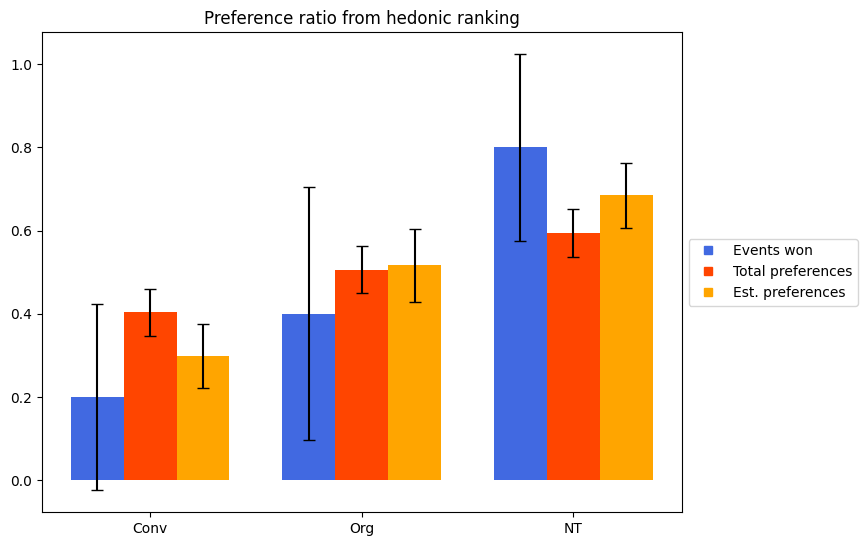}
\caption{Preference ratios for the whole dataset\label{fig3}}
\end{figure}

\begin{figure}[H]
\centering
\includegraphics[width=14 cm]{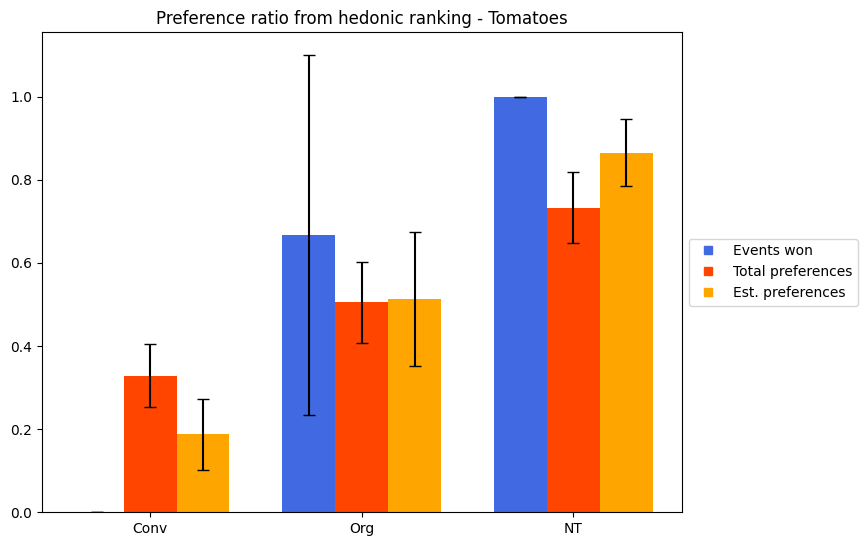}
\caption{Preference ratios when the tested vegetable is tomato\label{fig4}}
\end{figure}

Finally, we add some descriptive statistics (and confidence intervals) related with the whole dataset and the tomatoe special case. Figure \ref{fig3} and Figure \ref{fig4} compute the result of the preferences for each agricultural pratices, based on the 942 judges able to detect a difference (as well the correct answer not due to chance). For each modality among Conv, Org and NT, we count in Figure \ref{fig3} and Figure \ref{fig4} the number of success, in terms of events, as well as total preferences judge by judge. These statistics allow to illustrate a trend of growth from Conv to Org to NT. This trend is confirmed and becomes more significant in Figure \ref{fig4} with tomatoes. 
%\begin{figure}[H]
%\includegraphics[width=14 cm]{Definitions/correct_answers_ratio.png}
%\caption{Correct answers ratio by 2-uplet\label{fig2}}
%\end{figure}   
%\unskip

\unskip

\unskip

\section{Conclusion and perspectives}

We collect data from consumers to analyze sensory properties of seven fruits and vegetables based on different agricultural practices. Triangle test as well as hedonic ranking was repeated for each event for o total of 942 consumers. A statistical analysis was performed thanks to hypothesis testing and maximum likelihood estimation. The results illustrate globally significant p-values for differenciating two products from conventional, organic and no-till agricultural practices (10 out of 15 events). However, the results are less significant to show preferences thanks to hedonic ranking and testing procedure for each event.

Aggregation of the whole dataset reveals a trend Conv $<$ Org $<$ NT (P $< 0.2$). The special computation of tomatoes data also gives us particularly interesting results and both significant triangle and hedonic ranking comparing no-till and conventional practices, as well as no-till and organic techniques. Up to our knowledge, this is a first rigorous statistical hedonic ranking procedure that shows significant results between no-till versus organic techniques. It shows a first sensory signal for a preference of more sustainable products for these untrained consumers. An interesting future work could be to fully investigate tomatoes from these different practices in order to add more data and reach better significative results.

Another direction for future work could be to train panellists for the purpose of rating or comparing one vegetable to another, describing its sensory profile and detecting particular changes in the intensity of some specific sensory propertie, like odour, crunchyness, pronounced after-taste or aroma. Indeed, some consumers in our study reveal better tasting capacity, and give us relevant observations about the samples, and some particular properties, with respect to the origin, freshness and some tasting properties. From this side, a detailed study of tomatoes could be interesting to explain our preliminary results.

Finally, in this paper, we choose to compare different agricultural practices by selecting producers with conventional, organic or no-till techniques. This cutting is motivated by different soil managements like the prior loosening of the soil, soil decompaction, and other disturbances. Following \cite{morel2021}, it could be challenging to conduct a more in-depth qualitative investigation of our producers, in order to derive habits and concepts regarding farm labour and farm workers, and propose a critical examination of methods to better understand the main differences between organic and conventional farms, as well as the prominence of a better environmental soil management system. Indeed, even if we meet in this study a rather clear decomposition of practices in three main categories, frontier are often scarse.

%%%%%%%%%%%%%%%%%%%%%%%%%%%%%%%%%%%%%%%%%%
\vspace{6pt} 

%\paragraph{Funding} This research was partly funded by EMAC society, bipartite contrat with Université de Pau et des Pays de l'Adour, named 'Du sol à l'Assiette'.
\vspace{6pt}

\paragraph{Acknowledgments} We thank all the members of the team GreenAI UPPA, to help Florine and Sebastien to conduct the experiences, and particularly Nathan van Hoevelaken.

\bibliography{ref} 
\bibliographystyle{ieeetr}
\end{document}